\pdfoutput=1

\documentclass[11pt,twoside,a4paper,cmspaper,final,collab]{cms-tdr}
\def\svnVersion{ee096e3-D}\def\svnDate{2020/12/08}\def\cmsCernNoTag{CERN-EP-2020-127}\def\cmsCernDate{\today}\def\cmsMessage{"Published in Physics Letters B as \href{http://dx.doi.org/10.1016/j.physletb.2020.135992}{\doi{10.1016/j.physletb.2020.135992}.}"}

\begin{document}\cmsNoteHeader{SMP-20-001}

\hyphenation{had-ron-i-za-tion}
\hyphenation{cal-or-i-me-ter}
\hyphenation{de-vices}
\newlength\cmsFigWidth
\newlength\cmsTabSkip\setlength{\cmsTabSkip}{1ex}
\ifthenelse{\boolean{cms@external}}{\setlength\cmsFigWidth{\columnwidth}}{\setlength\cmsFigWidth{0.65\textwidth}}
\ifthenelse{\boolean{cms@external}}{\providecommand{\cmsLeft}{upper\xspace}}{\providecommand{\cmsLeft}{left\xspace}}
\ifthenelse{\boolean{cms@external}}{\providecommand{\cmsRight}{lower\xspace}}{\providecommand{\cmsRight}{right\xspace}}
\newcommand{\MG}{\textsc{MG5}\xspace} 
\newcommand{\ZZjj}{\ensuremath{\PZ\PZ\mathrm{jj}}\xspace} 
\newcommand{\ZZ}{\ensuremath{\PZ\PZ}\xspace} 
\newcommand{\mjj}{\ensuremath{m_{\mathrm{jj}}}\xspace}
\newcommand{\mfourl}{\ensuremath{m_{4\ell}}\xspace}
\newcommand{\DeltaEtajj}{\ensuremath{\Delta\eta_{\mathrm{jj}}}\xspace}

\cmsNoteHeader{SMP-20-001}

\title{Evidence for electroweak production of four charged leptons and two jets in proton-proton collisions at \texorpdfstring{$\sqrt{s} = 13\TeV$}{sqrt(s) = 13 TeV}}

\date{\today}

\abstract{Evidence is presented for the electroweak (EW) production of two jets (jj) in association with two \PZ bosons and constraints on anomalous quartic gauge couplings are set. The analysis is based on a data sample of proton-proton collisions at $\sqrt{s}=13\TeV$ collected with the CMS detector in 2016--2018, and corresponding to an integrated luminosity of 137\fbinv. The search is performed in the fully leptonic final state $\PZ\PZ\to\ell\ell\ell'\ell'$, where $\ell,\ell' = \Pe, \mu$. The EW production of two jets in association with two \PZ bosons is measured with an observed (expected) significance of 4.0 (3.5) standard deviations. The cross sections for the EW production are measured in three fiducial volumes and the result is $\sigma_{\mathrm{EW}}(\Pp\Pp\to \PZ\PZ\mathrm{jj}\to\ell\ell\ell'\ell'\mathrm{jj}) = 0.33 ^{+0.11}_{-0.10}\stat^{+0.04}_{-0.03}\syst\unit{fb}$ in the most inclusive volume, in agreement with the standard model prediction of $0.275 \pm 0.021\unit{fb}$. Measurements of total cross sections for jj production in association with two \PZ bosons are also reported. Limits on anomalous quartic gauge couplings are derived in terms of the effective field theory operators T0, T1, T2, T8, and T9.}

\hypersetup{%
pdfauthor={CMS Collaboration},%
pdftitle={Evidence for electroweak production of four charged leptons and two jets in proton-proton collisions at sqrt(s) = 13 TeV},%
pdfsubject={CMS},%
pdfkeywords={CMS, physics, SM, ZZ, VBS, aQGC}
}
\maketitle 

\section{Introduction}
\label{sec:introduction}

In the standard model (SM), the electroweak (EW) vector bosons, like the other fundamental particles, acquire their masses through the coupling to the Brout-Englert-Higgs field. The photon remains massless, with only two degrees of polarization (i.e., transverse), whereas the \PW and \PZ bosons
acquire an additional degree of freedom (i.e., longitudinal), as a consequence of the electroweak symmetry breaking (EWSB)~\cite{Higgs:1966ev, Englert:1964et}. Thus, the scattering of massive vector bosons is at the heart of the EWSB mechanism and its study can lead to significant insight into the origin of particle masses. Moreover, if the couplings between the Higgs boson and vector bosons ($\PH\PV\PV$) differ from their SM values, the subtle interplay between $\PH\PV\PV$, triple, and quartic gauge couplings as predicted in the SM is incomplete, and the cross section for the longitudinal scattering diverges at large scattering energies, eventually violating the unitarity.

\begin{figure*}[thb]
\centering
\includegraphics[width=0.65\textwidth]{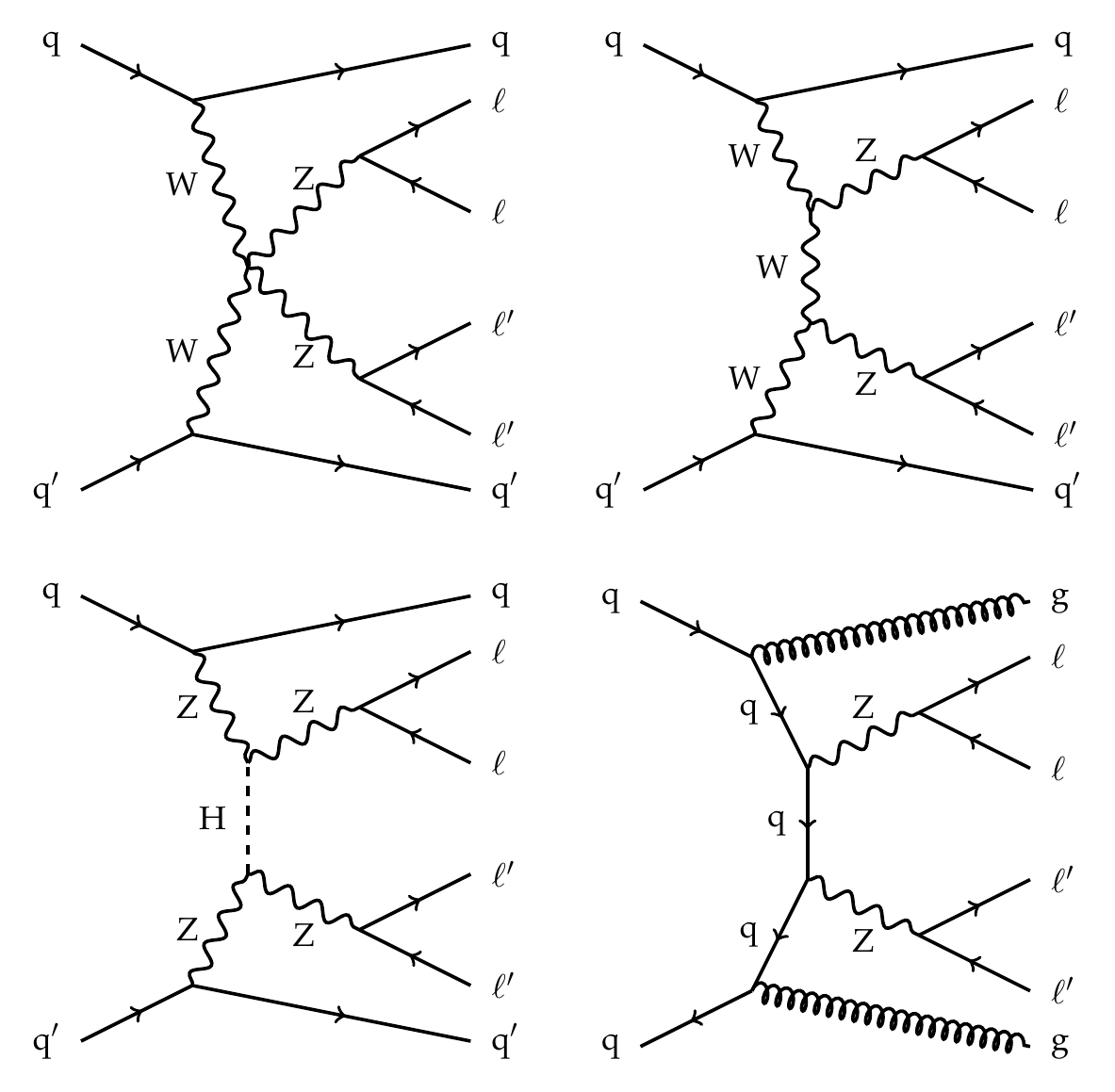}
\caption{Representative Feynman diagrams for the EW- (top row and bottom left) and QCD-induced (bottom right) production of the $\ZZjj\to\ell\ell\ell'\ell'\mathrm{jj}$ ($\ell,\ell' = \Pe, \mu$) final state. The scattering of massive gauge bosons as depicted in the top row is unitarized by the interference with amplitudes that feature the Higgs boson (bottom left).}
\label{fig:feynman}
\end{figure*}

At the CERN LHC, vector boson scattering (VBS) is the interaction of two EW vector bosons emitted by quarks ($\Pq$) from the two colliding protons. 
The VBS process is generally labeled by the type of outgoing vector bosons. The two jets (jj) originating from the scattered quarks are typically emitted in the forward-backward region of the detector, giving rise to events whose signature in the detector is characterized by a region in rapidity (so-called "rapidity gap")~\cite{rapidity_gap_1, rapidity_gap_2}, where no additional hadronic activity is expected from the hard scattering.
The decay of the vector bosons into fermions defines the final signature of the VBS-like event.
The pure VBS contributions, however, are embedded into a wider set of possible two-to-six processes, with which they interfere (Fig.~\ref{fig:feynman}).
All processes at the order of $\alpha_{\mathrm{EW}}^6$ (tree level) are considered as EW production (Fig.~\ref{fig:feynman} upper panels and bottom left panel), whereas the processes at the order $\alpha_{\mathrm{EW}}^4 \alpS^2$
where at tree level the jets are induced by quantum chromodynamics (QCD) (lower right panel in Fig.~\ref{fig:feynman}), constitute a background referred to as QCD-induced background. Kinematic requirements on the dijet system are used to define fiducial regions enriched in VBS-like events and where QCD-induced backgrounds are suppressed.

Both the ATLAS and CMS Collaborations have performed searches for the scattering of massive vector bosons,
using data from proton-proton (pp) collisions at the center-of-mass energy of 13\TeV.
The ATLAS Collaboration reported the observation of EW production of two jets in association with a same-sign \PW boson pair~\cite{ATLAS:ssWW}, with a $\PW\PZ$ boson pair~\cite{ATLAS:WZ}, and, recently, with a \PZ boson pair~\cite{ATLAS:ZZ}. Results were also reported on the measurement of the EW diboson production ($\PW\PW$, $\PW\PZ$, \ZZ) in association with a high-mass dijet system in semileptonic final states~\cite{ATLAS:VVsemilep}, with an observed significance of 2.7 standard deviations.
The CMS Collaboration observed the production of two EW-induced jets with two same-sign \PW bosons~\cite{CMS:ssWW,CMS:ssWWandWZ} and with $\PW\PZ$ pairs~\cite{CMS:ssWWandWZ}, and measured the EW production of jets in association with \ZZ~\cite{CMS:ZZ} with an observed significance of 2.7 standard deviations.

This paper presents evidence
for the EW production of two jets in association with two \PZ bosons,
where both \PZ bosons decay into electrons or muons, $\ZZ\to\ell\ell\ell'\ell'$ ($\ell,\ell' = \Pe, \mu$). Despite a low cross section, a small $\PZ \to \ell\ell$ branching fraction,
and a large QCD-induced background, this channel provides a clean leptonic final state with a small experimental background, where one or more
reconstructed lepton candidates originate from the misidentification of jet fragments or from nonprompt leptons.

The search for the EW-induced production of the $\ell\ell\ell'\ell' \mathrm{jj}$ final state is carried out using \Pp\Pp collisions at $\sqrt{s} = 13\TeV$ recorded with the CMS detector at the LHC.
The data set corresponds to an integrated luminosity of 137\fbinv collected in 2016, 2017, and 2018.  A discriminant based on
a matrix element likelihood approach (MELA)~\cite{Chatrchyan:2012xdj, Bolognesi:2012mm, Gao:2010qx,Anderson:2013afp,Gritsan:2016hjl} is used to extract the signal significance and to measure the cross sections for the EW and the EW+QCD production of the $\ell\ell\ell'\ell' \mathrm{jj}$ final state in a fiducial volume.
Finally, the selected $\ell\ell\ell'\ell' \mathrm{jj}$ events are used to constrain anomalous quartic gauge couplings (aQGC) described in the effective field theory approach~\cite{Degrande:2012wf} by the operators T0, T1, and T2, as well as the neutral-current operators T8 and T9~\cite{Eboli:2006wa}.

\section{The CMS detector}
\label{sec:cmsandselection}

The central feature of the CMS apparatus is a superconducting solenoid of 6\unit{m} internal diameter, providing a magnetic field of 3.8\unit{T}.
Within the solenoid volume are a silicon pixel and strip tracker, a lead tungstate crystal electromagnetic calorimeter, and a 
brass and scintillator hadron calorimeter, each composed of a barrel and two endcap sections. 
Forward calorimeters extend the pseudorapidity ($\eta$) coverage provided by the barrel and endcap detectors. 
Muons are detected in gas-ionization detectors embedded in the steel flux-return yoke outside the solenoid. 
The first level of the CMS trigger system, composed of custom hardware processors, uses information from the calorimeters and muon detectors to select events of interest with a latency of 3.2\mus. The high-level trigger processor farm further decreases the event rate from around 100\unit{kHz} to less than 1\unit{kHz}, before data storage~\cite{Khachatryan:2016bia}.
A more detailed description of the CMS detector, together with a definition of the coordinate system used and the relevant kinematic variables, 
can be found in Ref.~\cite{Chatrchyan:2008zzk}.

\section{Signal and background simulation}
\label{sec:mc}

Several Monte Carlo (MC) event generators are used to simulate signal and background contributions.
The simulated samples are employed to optimize the event selection, evaluate the signal efficiency and acceptance, and to model the signal and
irreducible background contributions in the signal extraction fit.

The EW production of two \PZ bosons and two final-state quarks, where the \PZ bosons decay leptonically, is simulated at leading order (LO) using 
\MGvATNLO~v2.4.6 (abbreviated as \MG in the following)~\cite{MGatNLO}. The leptonic $\PZ$ boson decays are simulated using \textsc{MadSpin}~\cite{Artoisenet:2012st}. The contribution of electrons and muons from $\tau$ decays to the signal is very small and is therefore neglected. The sample includes triboson processes, where the \PZ boson pair is accompanied by a third vector boson that decays hadronically, as well as diagrams involving the quartic gauge coupling vertex. The predictions from this sample are cross-checked with those obtained from the LO generator \textsc{Phantom}\ v1.2.8~\cite{Phantom} with agreement in the yields and the distributions exploited for the signal extraction.

The leading QCD-induced production
of two \PZ bosons in association with jets, whose contribution with two jets in the final state is referred to as $\qqbar\to\ZZjj$,	
is simulated at next-to-leading order (NLO) with 
\MG with up to two extra parton emissions, and merged with the parton shower simulation using the FxFx scheme~\cite{Frederix:2012ps}.
Next-to-next-to-leading order corrections calculated with MATRIX~v1.0.0~\cite{Grazzini2015407,Grazzini:2017mhc,Kallweit:2018nyv} are applied as $K$ factors, differentially as a function of the invariant mass of the \ZZ system ($m_{\ZZ})$. The resulting corrections range from 9\%, at values of $m_{\ZZ}$ close to 180 GeV, to 5\%, for high $m_{\ZZ}$ values.  Additional NLO EW corrections are applied for $m_{\ZZ} > 2m_{\PZ}$, following the calculations from Ref.~\cite{Gieseke:2014gka}. These corrections become larger with increasing values of $m_{\ZZ}$ and are below 5\% for $m_{\ZZ} < 600$ GeV.

The interference between the EW and QCD diagrams is evaluated using dedicated samples produced with \MG at LO, via the direct generation of the interference term between the two processes.
 
The loop-induced production of two \PZ bosons from a gluon-gluon ($\Pg\Pg$) initial state, whose contribution with two jets in the final state is referred to as $\Pg\Pg \to\ZZjj$, is simulated at LO with up to two extra parton emissions using \MG by explicitly requiring a loop-induced process~\cite{Li:2020nmi}. For the 1- and 2-jet contributions, a $\Pp\Pp$ initial state instead of $\Pg\Pg$ is specified in \MG to also include initial-state radiation contributions where a gluon involved in the hard process is emitted from an initial quark. Finally, the samples with 0 to 2 extra partons are merged with parton shower simulation using the MLM matching scheme~\cite{MLM,MLM2}. An NLO/LO $K$ factor, which is extracted from Refs.~\cite{Caola:2015psa,Caola:2016trd}, is used to normalize this process.

Background processes that contain four prompt, isolated leptons and additional jets in the
final state, namely $\ttbar\PZ$ and $\PV\PV\PZ$ ($\PV = \PW,\PZ$), are simulated with \MG at NLO.  

The simulation of the aQGC processes is performed at LO using \MG and employs matrix element reweighting to obtain a finely spaced grid for each of the five anomalous couplings probed by the analysis.

The \PYTHIA 8.226 and 8.230~\cite{Sjostrand:2015} package versions
are used for parton showering, hadronization and
the underlying event simulation, with parameters set by the CUETP8M1
tune~\cite{Khachatryan:2015pea} (CP5 tune~\cite{CP5}) for the 2016 (2017 and 2018) data-taking period. The NNPDF3.0 (NNPDF3.1) set of
parton distribution functions, PDFs~\cite{NNPDF2015}, is used for the 2016 (2017 and 2018) data-taking period.
Unless specified otherwise, the simulated samples are normalized 
to the cross sections obtained from the respective event generator.

The detector response is simulated using a detailed
description of the CMS detector implemented in the \GEANTfour
package~\cite{GEANT, Geant2}. The simulated events are  reconstructed
using the same algorithms used for the data, 
and include additional interactions in the same and neighboring bunch crossings, referred to as pileup.
Simulated events are weighted so that the pileup distribution reproduces that observed in the data, which has an average of about 23 (32) interactions per bunch crossing in 2016 (2017 and 2018).

\section{Event reconstruction and selection}
\label{sec:eventselection}

The final state consists of at least two pairs of oppositely charged isolated leptons and at least two hadronic jets. The \ZZ selection is similar to that used in the CMS $\PH\to\ZZ\to\ell\ell\ell^{\prime}\!\ell^{\prime}$ measurement~\cite{Sirunyan_2017}. 

The primary triggers require the presence of a
pair of loosely isolated leptons, whose exact requirements depend on the data-taking year. Triggers requiring three leptons with low transverse momentum (\pt), as well as isolated single-electron and single-muon triggers, help to recover efficiency. 
The overall trigger efficiency for events that satisfy the \ZZ selection described below is $> 98\%$.

Events are reconstructed using a particle-flow algorithm~\cite{CMS-PRF-14-001} that identifies each individual particle with an optimized combination of all subdetector information. 
The candidate vertex with the largest value of summed physics-object $\pt^2$ is the primary $\Pp\Pp$ interaction vertex. The physics objects are the jets, clustered using the jet finding algorithm~\cite{Cacciari:2008gp,Cacciari:2011ma} with the tracks assigned to candidate vertices as inputs, and the associated missing transverse momentum (\ptmiss), taken as the negative vector sum of the \pt of those jets (which include the leptons). 

Electrons are identified using a multivariate classifier, which includes observables sensitive to 
bremsstrahlung along the electron trajectory, the geometrical and energy-momentum compatibility between the 
electron track and the associated energy cluster in the electromagnetic calorimeter, the shape of the electromagnetic shower, isolation variables,
and variables that discriminate against electrons originating from photon conversions~\cite{Khachatryan:2015hwa}. 

Muons are reconstructed by combining 
information from the silicon tracker and the muon system~\cite{Sirunyan:2018fpa}.
The matching between the muon-system and tracker tracks proceeds either outside-in, starting from a track in the muon system, 
or inside-out, starting from a track in the silicon tracker. 
The muons are selected from the reconstructed muon track candidates by applying minimal requirements on the track in both 
the muon system and silicon tracker.

To further suppress electrons from photon conversions and muons originating from in-flight decays of hadrons, the three-dimensional impact parameter of each lepton track, computed with respect to the primary vertex position, is required to be less than four times the uncertainty in the impact parameter.

Leptons are required to be isolated from other particles in the event. The
relative isolation is defined as
\begin{linenomath}
\begin{equation}
R_\text{iso} = \bigg[ \sum_{\substack{\text{charged} \\ \text{hadrons}}} \!\! \pt \, + \,
\max\big(0, \sum_{\substack{\text{neutral} \\ \text{hadrons}}} \!\! \pt
\, + \, \sum_{\text{photons}} \!\! \pt \, - \, \pt^\mathrm{PU}
\big)\bigg] \bigg/ \pt^{\ell},
\label{eq:iso}
\end{equation}
\end{linenomath}
where the scalar sums run over the charged and neutral hadrons, as well as the photons, in a
cone defined by
$\Delta R \equiv \sqrt{\smash[b]{\left(\Delta\eta\right)^2 + \left(\Delta\phi\right)^2}} = 0.3$
around the lepton trajectory, where $\eta$ and $\phi$ denote the azimuthal angle and pseudorapidity of the particle, respectively.
To minimize the contribution of charged particles from pileup to the isolation calculation,
charged hadrons are included only if they originate from the
primary vertex. The contribution of
neutral particles from pileup $\pt^\mathrm{PU}$
is evaluated for electrons with the jet area method described in Ref.~\cite{Cacciari:2007fd}.
For muons, $\pt^\mathrm{PU}$ is taken as half the \pt sum of all charged particles in the cone originating
from pileup vertices. The factor of one-half accounts for the expected ratio of charged to neutral particle production in hadronic interactions.
Muons with $R_\text{iso} < 0.35$ are considered isolated, whereas for electrons, the $R_\text{iso}$ variable is included in the multivariate classifier.

The lepton reconstruction and selection efficiency is 
measured in bins of $\pt^\ell$ 
and $\eta^\ell$ using the tag-and-probe technique~\cite{Khachatryan_2011} on events with single \PZ bosons. The measured efficiencies are used to correct the simulation.
The muon (electron) momentum scales are calibrated in bins of $\pt^\ell$ and $\eta^\ell$ 
using the \PJGy meson and \PZ boson (\PZ boson only) leptonic decays.

Jets are reconstructed from particle-flow candidates using the anti-\kt clustering algorithm~\cite{Cacciari:2008gp}, as implemented in the \FASTJET package~\cite{Cacciari:2011ma}, with a distance parameter of 0.4. To ensure a good reconstruction efficiency and to reduce the instrumental background, as well as the contamination from pileup, loose identification criteria based on the multiplicities and energy fractions carried by charged and neutral hadrons are imposed on jets~\cite{CMS-PAS-JME-16-003}. Only jets with $\abs{\eta} < 4.7$ are considered.

Jet energy corrections are extracted from data and simulated events to account for the effects of pileup, uniformity of the detector response, and residual differences between the jet energy scale in data and simulation.
The jet energy scale calibration~\cite{1748-0221-6-11-P11002, Khachatryan:2016kdb} relies on corrections parameterized in terms of the uncorrected \pt and $\eta$ of the jet, and is applied as a multiplicative factor, scaling the four-momentum vector of each jet. To ensure that jets are well measured and to reduce the pileup contamination, all jets must have a corrected $\pt >30\GeV$. Jets from pileup are further rejected using pileup jet identification criteria based on the compatibility of the associated tracks with the primary vertex inside the tracker acceptance and on the topology of the jet shape in the forward region~\cite{JETPU}.

A signal event must contain at least two
$\PZ$ candidates, each formed from pairs of isolated electrons or muons of opposite charges.
Only reconstructed electrons (muons) with $\pt > 7~(5) \GeV$ are considered.
At least two leptons are required to have $\pt > 10 \GeV$ and at least one is required to have $\pt > 20 \GeV$.
All leptons are required to be separated by
$\Delta R \left(\ell_1, \ell_2 \right) > 0.02$, and electrons are required to be separated from muons by
$\Delta R \left(\Pe, \mu \right) > 0.05$.

Within each event, all permutations of leptons giving a valid pair of $\PZ$
candidates are considered. For each \ZZ candidate, the lepton pair with the invariant mass closest to the nominal \PZ boson mass is denoted $\PZ_1$.
The other dilepton candidate is denoted $\PZ_2$. Both $m_{\PZ_1}$ and $m_{\PZ_2}$ are required to be in the range 60--120\GeV.
All pairs of oppositely charged leptons that can be built from the \ZZ candidate, regardless of flavor, are required to
satisfy ${m_{\ell \ell^{\prime}\!} > 4\GeV}$ to suppress backgrounds from hadron decays. If multiple \ZZ candidates in an event pass this selection, the one with the largest scalar \pt sum of the $\PZ_2$ leptons is retained.
Finally, the invariant mass of the four leptons is required to satisfy $\mfourl > 180 \GeV$. 
This selection is referred to as the \ZZ selection.

The search for the EW production of two $\PZ$ bosons is performed on a 
subset of events that pass the \ZZ selection, namely those with at least two jets. The jets are required to be separated from the leptons of the \ZZ candidate by $\Delta R>0.4$. The two highest \pt jets 
are referred to as the tagging jets and their invariant mass (\mjj) is 
required to be $>100\GeV$. This selection is referred to as the 
\ZZjj inclusive selection and is used to measure the signal significance, the total fiducial cross sections, and to perform the aQGC search. Additionally, two VBS signal subregions are defined for fiducial cross section measurements in signal-enriched regions: a loose VBS signal-enriched region that requires $\mjj>400\GeV$ and $\abs{\DeltaEtajj}>2.4$ and corresponds to a signal purity of 
${\approx}20\%$, and a tight VBS signal-enriched region that requires
${\mjj>1\TeV}$ and $\abs{\DeltaEtajj}>2.4$ 
and corresponds to a signal purity of 
${\approx}50\%$. Finally, a background control region is defined from events that satisfy the \ZZjj inclusive selection but fail at least one of the criteria that define the loose VBS signal-enriched region.

\section{Background estimation}
\label{sec:bckgestimation}

The dominant background arises from the production of two \PZ bosons in association with QCD-induced jets.
The yield and shape of the matrix element discriminant for this irreducible background are taken from simulation, but ultimately constrained by the data in the fit that extracts the EW signal, as described in Section~\ref{sec:VBS}. Other irreducible backgrounds arise from processes that produce four genuine 
high-\pt isolated leptons, $\Pp\Pp\to\ttbar\PZ\mathrm{+jets}$ and $\Pp\Pp\to\PV\PV\PZ\mathrm{+jets}$.
These small contributions feature kinematic distributions similar to that of the dominant background and are estimated from simulation.

Reducible backgrounds arise from processes in which heavy-flavor jets produce 
secondary leptons or from processes in which jets are misidentified as leptons. They are referred to as $\PZ\mathrm{+X}$ and are predominately composed of $\PZ\mathrm{+jets}$ events, with minor contributions from $\ttbar\mathrm{+jets}$ and $\PW\PZ\mathrm{+jets}$ processes.
The lepton identification and isolation requirements significantly suppress this
background, which is only 2--3\% after the \ZZjj inclusive selection and is even smaller in the signal region. 
This reducible contribution is estimated from data by weighting events from a control region by a lepton misidentification rate, which is also determined from data.  
Events in the control region satisfy the \ZZjj inclusive selection, with the exception that the $\PZ_2$ is composed of same-sign same-flavor leptons (SS-SF). The SS-SF leptons are required to originate from the primary vertex without any identification or isolation requirement.

The lepton misidentification rate is measured by selecting events that feature one \PZ boson candidate and a third reconstructed lepton. 
The fraction of events for which the third lepton satisfies the identification and isolation criteria is the lepton misidentification rate. The misidentification rates are evaluated using the tight requirement $\vert m_{\PZ_1} - m_{\PZ} \vert < 7 \GeV$ to reduce the contribution from asymmetric photon conversions, and $\ptmiss < 25 \GeV$ to suppress the $\PW\PZ$ contribution. 
 
We validate the procedure using a second control region from opposite-sign same-flavor leptons that fail the selection criteria. The procedure is identical to that used in Ref.~\cite{Sirunyan_2017}.

\section{Systematic uncertainties}
\label{sec:syst}

The uncertainties in the QCD renormalization and factorization scales for the signal and in the jet energy scale 
are the two dominant systematic uncertainties in the measurement. The impact of the variation from each source of uncertainty is summarized below. All quoted ranges correspond to variations for the different leptonic final states and fiducial analysis regions.

Renormalization and factorization scale uncertainties are evaluated by varying both scales independently. The following variations from the default scale choice $\mu_{\mathrm{R}} = \mu_{\mathrm{F}} \equiv \mu_0$ are considered: $[\mu_{\mathrm{F}}, \mu_{\mathrm{R}}] = [\mu_0,\mu_0/2]$, $[\mu_0,2\mu_0]$, $[\mu_0/2,\mu_0]$, $[2\mu_0,\mu_0]$, $[\mu_0/2,\mu_0/2]$, $[2\mu_0,2\mu_0]$, taking the largest variation as the systematic uncertainty, which 
is about 6\% for the EW signal, 11\% for the interference term, and ranges from 10 to 12\% for the $\qqbar\to\ZZjj$ QCD background, which is described at a higher QCD order. 

Since the uncertainty in $\Pg\Pg\to\ZZjj$ that relates to missing higher order corrections are accounted for using a $K$ factor, an uncertainty in the normalization of 11\% is used, as derived from 
Refs.~\cite{Caola:2015psa,Caola:2016trd}.
The PDF and related $\alpS$ variations are evaluated 
from the variations of the respective eigenvalues set following the NNPDF prescription~\cite{NNPDF2015}, and are 3.2$\%$ (6.6$\%$) for the $\qqbar\to\ZZjj$ QCD background (EW signal). Although the PDFs used are different in the various years (see Section~\ref{sec:mc}), the associated uncertainties are very similar. Given the small dependence on the discriminant value, a constant value of 3--6\% is used for these uncertainties, depending on the sample considered.

Although in all simulated samples additional partons are described at the LO in matrix-elements or better, we investigate residual uncertainties from parton-shower modeling. Following the prescription from Ref.~\cite{Mrenna:2016sih}, the renormalization scales are varied independently for the initial- and final-state radiations by factors of 0.5 and 2, and alternative samples are simulated using \HERWIG 7~\cite{Bellm:2015jjp} with the CH3 tune~\cite{PASGEN19}. The largest deviation from the nominal value is used as the uncertainty. On average it ranges from 4\%, for the $\Pg\Pg\to\ZZjj$ background and EW signal, to 5\% for $\qqbar\to\ZZjj$, and is up to 16\% at the lowest values of $K_\mathrm{D}$.

The impact of the jet energy scale uncertainty ranges from 4.9 to 11.4\% (0.7 to 1.2\%) for the $\qqbar\to\ZZjj$ QCD background (EW signal) and the impact of the jet energy resolution uncertainty~\cite{Khachatryan:2016kdb} is 2.2--6.3\% (0.2--0.4\%).
The uncertainty in the trigger as well as the lepton reconstruction and selection efficiencies ranges from 2.5--9\%.
The uncertainty in the integrated luminosity is 2.3--2.5\% depending on the data-taking period~\cite{CMS-PAS-LUM-17-001,CMS-PAS-LUM-17-004,CMS-PAS-LUM-18-002}. The uncertainty in the estimate of the reducible background from control samples ranges from 33\% to 45\%, depending on the final state. This uncertainty includes the limited number of events
in the control regions as well as differences in background composition between the control regions used to determine the lepton misidentification rates and those used to estimate the yield in the signal region. The uncertainty from the limited size of the MC samples amounts to 2.5--4.2\% for the $\qqbar\to\ZZjj$ QCD background, 3.2\% for the $\Pg\Pg\to\ZZjj$ QCD background, and is $<1\%$ for the EW signal.
For $\ttbar\PZ$ and $\PV\PV\PZ$, the limited MC sample size is the dominant source of uncertainty, ranging from 19 to 24\%, while theory uncertainties range 9--12\%.

\section{Search for the EW production of \texorpdfstring{$\PZ\PZ$}{ZZ} with two jets}
\label{sec:VBS}

After the \ZZjj inclusive selection, the expected EW signal purity is about 6\% with 85\% of events coming from the QCD-induced production.
Additional kinematic selections are therefore necessary to enhance the contribution from EW production.  Table~\ref{tab:res_yields} presents the expected
and observed event yields for the \ZZjj inclusive selection, as well as for the loose and tight VBS signal-enriched selections.

\begin{table*}[!th]
\centering
\topcaption{Predicted signal and background yields with total uncertainties, and observed number of events for the \ZZjj inclusive selection and for the VBS loose and tight signal-enriched selections. Integrated luminosities per data set are reported in parentheses.}
\resizebox{\textwidth}{!}{
   \begin{tabular}{l c c c c c c c}
   \hline 
	Year		&  Signal (EW \ZZjj)	& Z+X	&  $\qqbar\to\ZZjj$	& $\Pg\Pg\to\ZZjj$ & $\ttbar\PZ$+$\PV\PV\PZ$ & Total predicted & Data\\
    \hline 
	& & & \ZZjj inclusive & & & & \\ [\cmsTabSkip]

  2016 (36\fbinv) &  6.3 $\pm$ 0.7 & 2.8 $\pm$ 1.1 & 65.6 $\pm$ 9.5 & 13.5 $\pm$ 2.0 & 8.4 $\pm$ 2.2 & 96 $\pm$ 13 & 95 \\ 
 2017 (41\fbinv) &   7.4 $\pm$ 0.8 & 2.4 $\pm$ 0.9 & 77.7 $\pm$ 11.2 & 20.3 $\pm$ 3.0 & 9.6 $\pm$ 2.5 & 117 $\pm$ 15 &  111 \\
 2018 (60\fbinv) &   10.4 $\pm$ 1.1 & 4.1 $\pm$ 1.6 & 98.1 $\pm$ 14.2 & 29.1 $\pm$ 4.3 & 14.2 $\pm$ 3.8 & 156 $\pm$ 20 & 159  \\
 All (137\fbinv) &    24.1 $\pm$ 2.5 & 9.4 $\pm$ 3.6 & 241.5 $\pm$ 34.9 & 62.9 $\pm$ 9.3 & 32.2 $\pm$ 8.5 & 370 $\pm$ 48 & 365  \\ [\cmsTabSkip]    
        & & & VBS signal-enriched (loose)& & & & \\ [\cmsTabSkip]

 2016 (36\fbinv) & 4.2 $\pm$ 0.4 & 0.4 $\pm$ 0.2 & 9.7 $\pm$ 1.4 & 3.2 $\pm$ 0.5 & 1.1 $\pm$ 0.3 & 18.7 $\pm$ 2.3 & 21 \\ 
2017 (41\fbinv) &  4.9 $\pm$ 0.5 & 0.5 $\pm$ 0.2 & 13.5 $\pm$ 1.9 & 5.5 $\pm$ 0.8 & 1.2 $\pm$ 0.3 & 25.5 $\pm$ 3.1 & 17  \\
2018 (60\fbinv) &  6.9 $\pm$ 0.7 & 0.8 $\pm$ 0.3 & 14.9 $\pm$ 2.2 & 8.3 $\pm$ 1.2 & 1.7 $\pm$ 0.5 & 32.6 $\pm$ 3.9 & 30   \\
All (137\fbinv) &   16.0 $\pm$ 1.7 & 1.6 $\pm$ 0.6 & 38.1 $\pm$ 5.5 & 17.0 $\pm$ 2.5 & 4.1 $\pm$ 1.1 & 76.8 $\pm$ 9.3 & 68  \\ [\cmsTabSkip]    
        & & & VBS signal-enriched (tight)& & & & \\ [\cmsTabSkip]

 2016  (36\fbinv) & 2.4 $\pm$ 0.3 & 0.10 $\pm$ 0.04 & 1.3 $\pm$ 0.2 & 0.7 $\pm$ 0.1 & 0.24 $\pm$ 0.06 & 4.8 $\pm$ 0.5 & 4\\
 2017 (41\fbinv) & 2.7 $\pm$ 0.3 & 0.05 $\pm$ 0.02 & 1.9 $\pm$ 0.3 & 1.2 $\pm$ 0.2 & 0.14 $\pm$ 0.04 & 6.0 $\pm$ 0.7 & 3\\
 2018 (60\fbinv) & 3.9 $\pm$ 0.4 & 0.17 $\pm$ 0.06 & 2.0 $\pm$ 0.3 & 1.5 $\pm$ 0.2 & 0.30 $\pm$ 0.08 & 7.8 $\pm$ 0.9 & 10\\
 All (137\fbinv) & 9.0 $\pm$ 1.0 & 0.32 $\pm$ 0.12 & 5.3 $\pm$ 0.8 & 3.3 $\pm$ 0.5 & 0.68 $\pm$ 0.18 & 18.6 $\pm$ 2.1 & 17 \\
    \hline 
   
   \end{tabular}
   }
      \label{tab:res_yields}
\end{table*}

The determination of the signal strength for the EW production, i.e., the ratio of the measured cross section to the SM expectation $\mu=\sigma/\sigma_{\mathrm{SM}}$, utilizes a matrix element discriminant ($K_\mathrm{D}$) to separate the signal and the QCD background. The discriminant is constructed following the approach described in Refs.~\cite{Gao:2010qx,Bolognesi:2012mm,Anderson:2013afp}: it utilizes matrix element calculations for the EW \ZZjj and $\qqbar\to\ZZjj$ processes from \MCFM~\cite{MCFM} and employs both the kinematical distributions of leptons and jets to separate signal from background.

The performance of the $K_\mathrm{D}$ discriminant was checked against a multivariate discriminant based on a boosted decision tree (BDT) employing seven input variables (\mjj, \DeltaEtajj, \mfourl, $\eta_{\PZ_{1}}^{*}$, $\eta_{\PZ_{2}}^{*}$, $R(\pt^{\mathrm{hard}})$, $R(\pt^{\mathrm{jets}})$) as defined and used in Ref.~\cite{CMS:ZZ}. Furthermore,
a BDT using up to 28 input variables, including the above as well as those used in Ref.~\cite{ATLAS:ZZ}, was studied and no significant gain was obtained.
This confirms that the $K_\mathrm{D}$ discriminant captures the differences between the kinematical distributions of signal and background events.

\begin{figure*}[!htbp]
\centering
\includegraphics[width=0.45\textwidth]{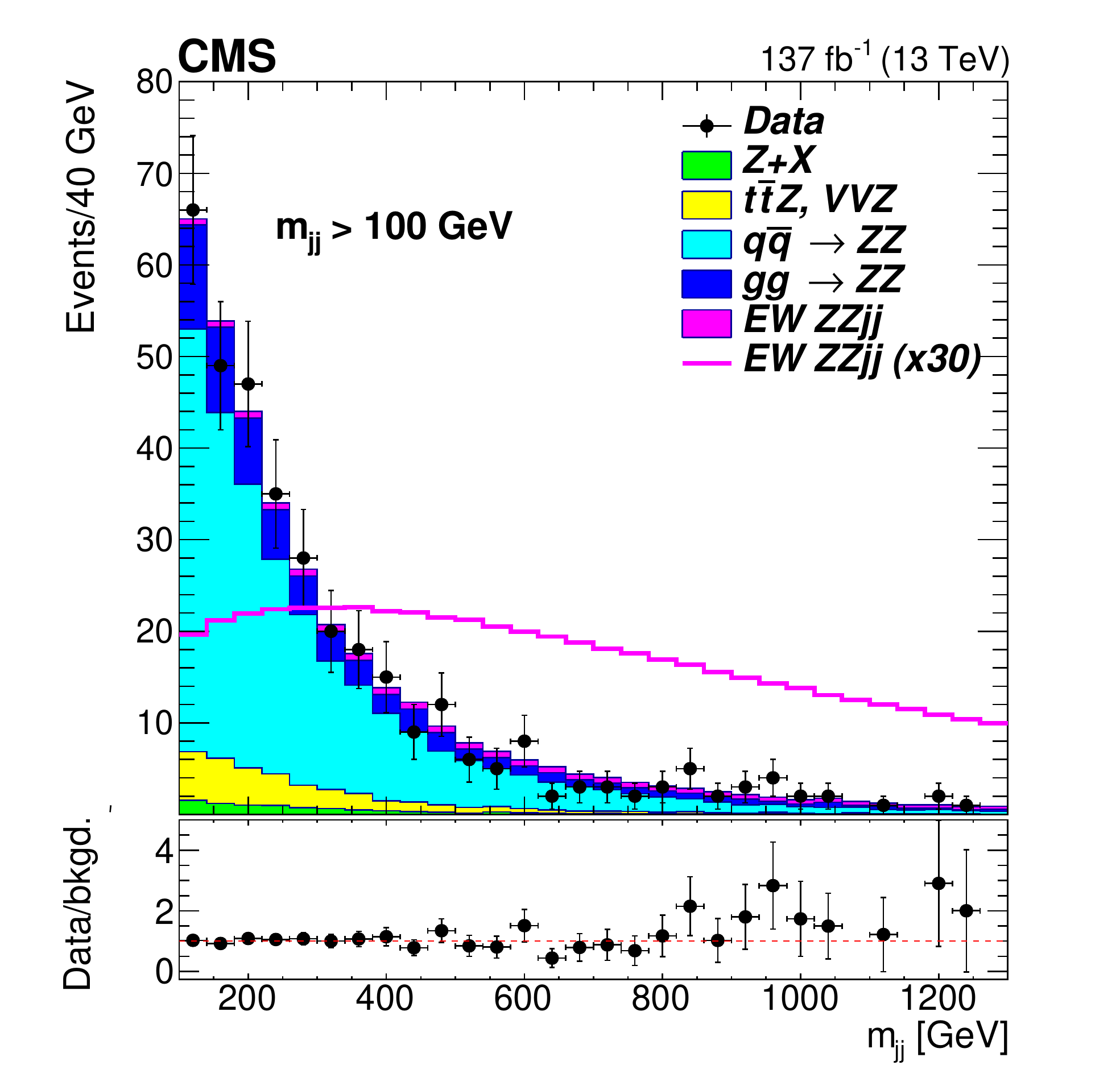}
\includegraphics[width=0.45\textwidth]{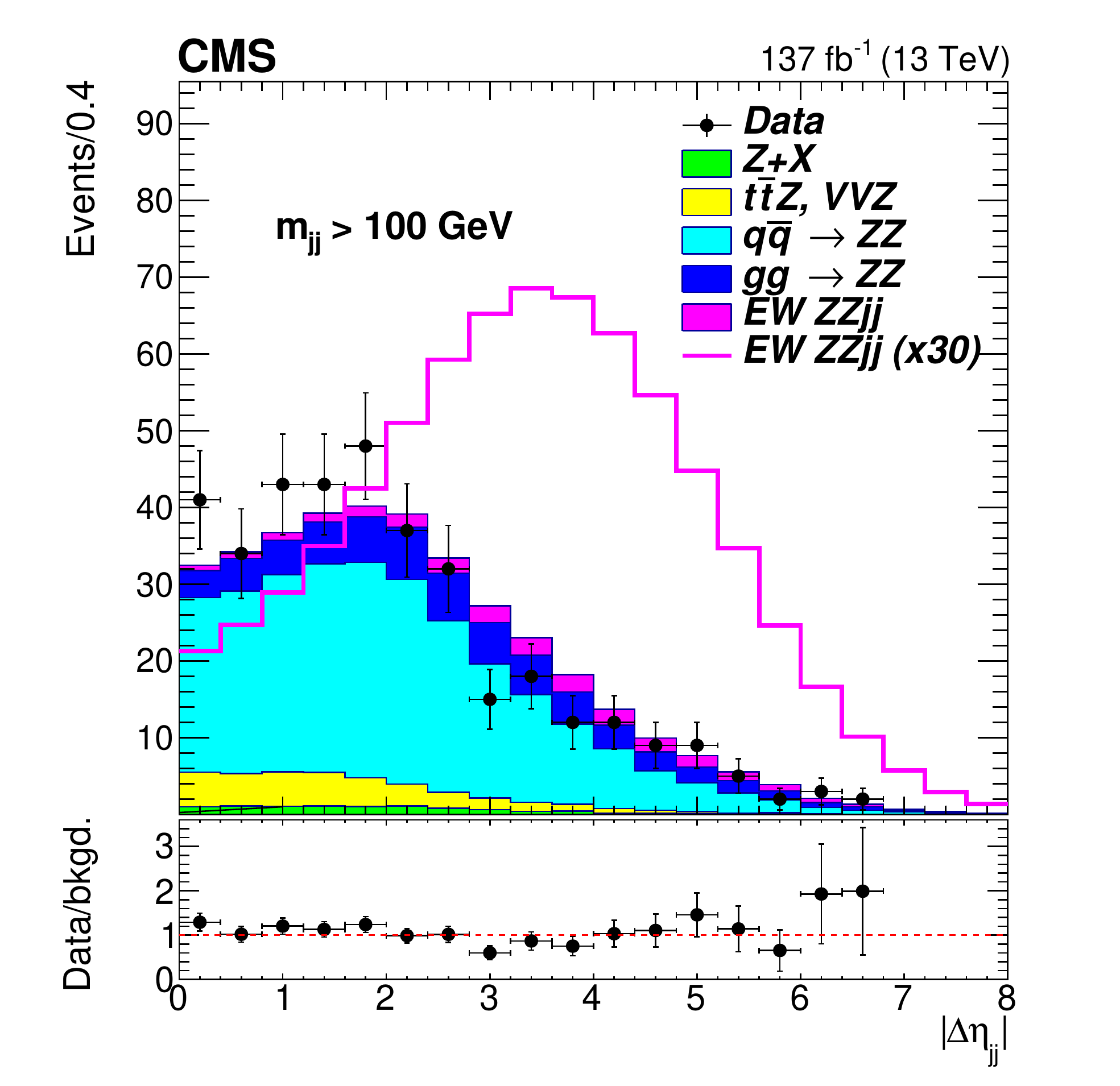}
\caption{
Distribution of \mjj (left) and $\abs{\DeltaEtajj}$ (right)
for events satisfying the \ZZjj inclusive selection. Points represent the data, filled histograms the expected signal and background contributions (stacked). The unfilled purple histograms represent the EW contribution (not stacked), scaled by a factor of 30. The lower panels show the ratio of the number of events in the data to the total number of expected background events.}
\label{fig:res_data_mc}
\end{figure*}

\begin{figure}[!htb]
\centering
\includegraphics[width=\cmsFigWidth]{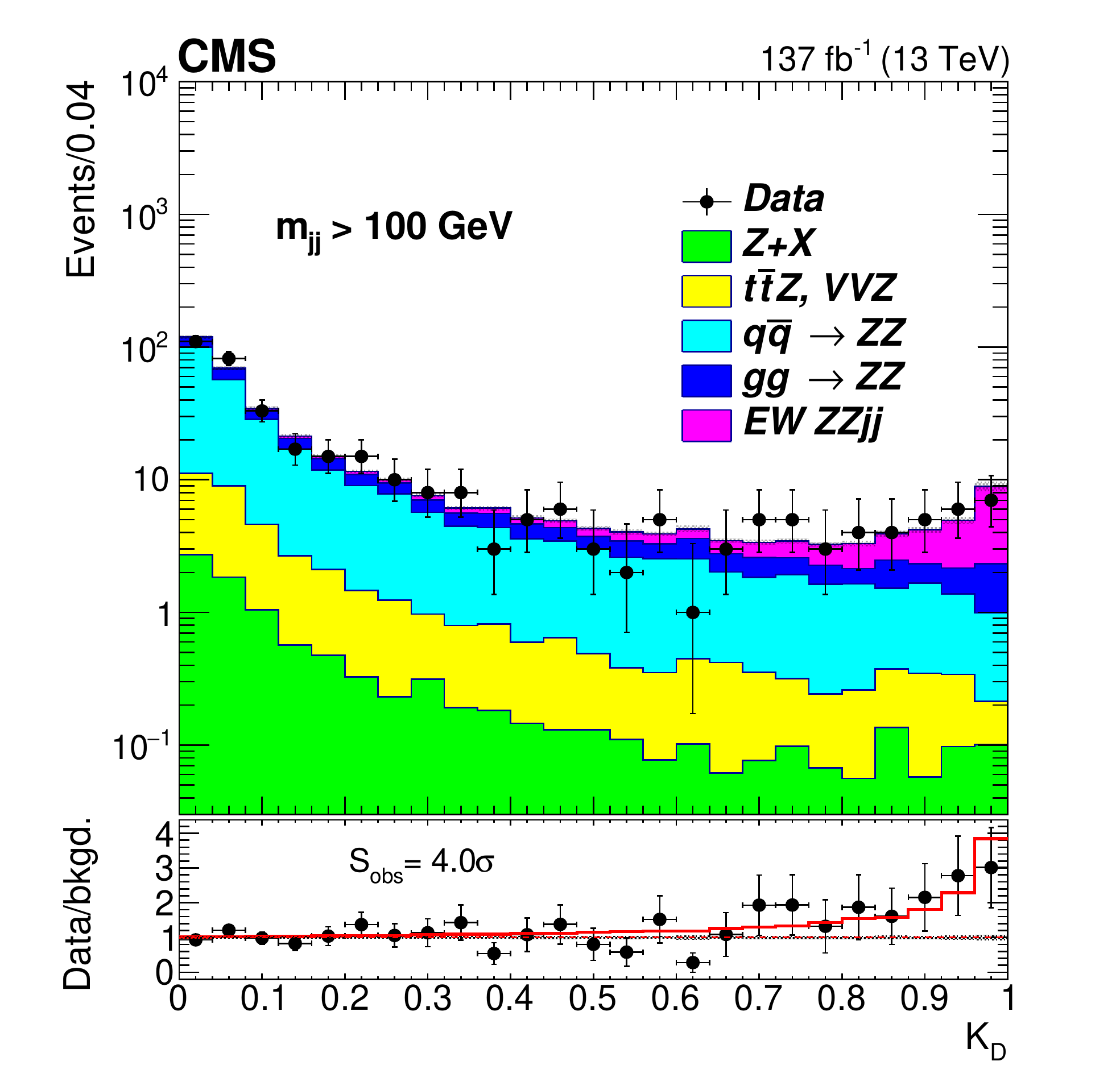}
\caption{
Postfit distributions of the matrix element discriminant  
for events satisfying the \ZZjj inclusive selection. Points represent the data, filled histograms the fitted signal and background contributions. The gray bands represent the uncertainties obtained from the fit covariance matrix. In the lower panel, points show the ratio of the number of events in the data to the total number of background events, with the red line indicating the ratio of the fitted total distribution to its background-only component. 
The observed significance is indicated in the lower panel.}
\label{fig:res_mela}
\end{figure}

Figure~\ref{fig:res_data_mc} presents the \mjj and $\abs{\DeltaEtajj}$ distributions in the \ZZjj inclusive region. 
The distribution of the $K_\mathrm{D}$ discriminant for all events in the \ZZjj inclusive selection is shown in Fig.~\ref{fig:res_mela}. The high signal purity contribution is visible at large discriminant values.

The distribution of the $K_\mathrm{D}$ discriminant for the backgrounds is validated in the background control region defined by selecting events with $\mjj<400\GeV$ or $\abs{\DeltaEtajj}<2.4$. A good agreement is observed between the data and the SM expectation.

The $K_\mathrm{D}$ discriminant distribution for events in the \ZZjj inclusive selection is used to extract the significance and signal strength of the EW signal via a maximum-likelihood fit. The expected distributions 
for the signal and the irreducible backgrounds are taken from the simulation while 
the reducible background is estimated from the data.
The shape and normalization of each distribution are allowed to vary in the fit within the respective uncertainties.
This approach constrains the yield of the QCD-induced production from the background-dominated region of the discriminant distribution.  
The signal strength of the EW signal in the \ZZjj inclusive selection is also determined from the same fit. Separate fits are used to determine the EW signal strengths in the other two analysis regions. Fits that only use the event counts in the three regions are performed to determine the signal strengths of the EW+QCD \ZZjj production. 

The systematic uncertainties in shape and normalization are treated as nuisance parameters in the fits and profiled \cite{ATLAS:2011tau}. The size of the interference between the EW and QCD production is very small (9\% and 3.5\% of the EW signal in the \ZZjj inclusive and VBS-enriched tight region, respectively). Its effect is included in the EW signal fits via a square-root scaling of the signal strength, approximated with a linear expansion to simplify the fitting technique, while it is neglected in the EW+QCD fits.

The measured signal strengths from the fits are used to determine the fiducial cross sections for the EW and the EW+QCD production. The fiducial volumes are almost identical to the selections imposed at the reconstruction level, and are detailed in Table~\ref{tab:fid}.
The generator-level lepton momenta are corrected by adding the momenta of generator-level photons within $\Delta R(\ell, \gamma) < 0.1$. The kinematic requirements to select \PZ boson candidates and the final \ZZjj candidate are the same as those used for the reconstruction-level analysis.

\begin{table*}[!htb]
\centering
\topcaption{Particle-level selections used to define the fiducial regions for EW and EW+QCD cross sections.}
   \begin{tabular}{l l}
   \hline 
	Particle type & \multicolumn{1}{c}{Selection} \\
    \hline 
	 \multicolumn{2}{c}{\ZZjj inclusive}  \\ [\cmsTabSkip] 
 Leptons & $\pt(\ell_1) > 20\GeV$ \\
         & $\pt(\ell_2) > 10\GeV$ \\
         & $\pt(\ell) > 5\GeV$ \\
         & $\abs{\eta(\ell)} < 2.5$ \\
 \PZ and \ZZ & $60 < m(\ell\ell) < 120\GeV$ \\
          & $m(4\ell) > 180 \GeV$      \\ [\cmsTabSkip]
 Jets     & at least 2   \\
          & $\pt(\mathrm{j}) > 30\GeV$   \\
          & $\abs{\eta(\mathrm{j})} < 4.7$ \\
          & $\mjj > 100\GeV$ \\
          & $\Delta R(\ell,\mathrm{j}) > 0.4$ for each $\ell, \mathrm{j}$ \\ [\cmsTabSkip]

          \multicolumn{2}{c}{VBS-enriched (loose)}  \\ [\cmsTabSkip]   
          & \multicolumn{1}{l}{\ZZjj inclusive +}  \\
          Jets     & $\abs{\DeltaEtajj} > 2.4$   \\
          & $\mjj > 400\GeV$ \\[\cmsTabSkip]

          \multicolumn{2}{c}{VBS-enriched (tight)}  \\ [\cmsTabSkip]   
          & \multicolumn{1}{l}{\ZZjj inclusive +}  \\
          Jets     & $\abs{\DeltaEtajj} > 2.4$   \\
          & $\mjj > 1\TeV$ \\
\hline      
   \end{tabular}
      \label{tab:fid}
\end{table*}

\begin{figure}[thb]
\centering
\includegraphics[width=\cmsFigWidth]{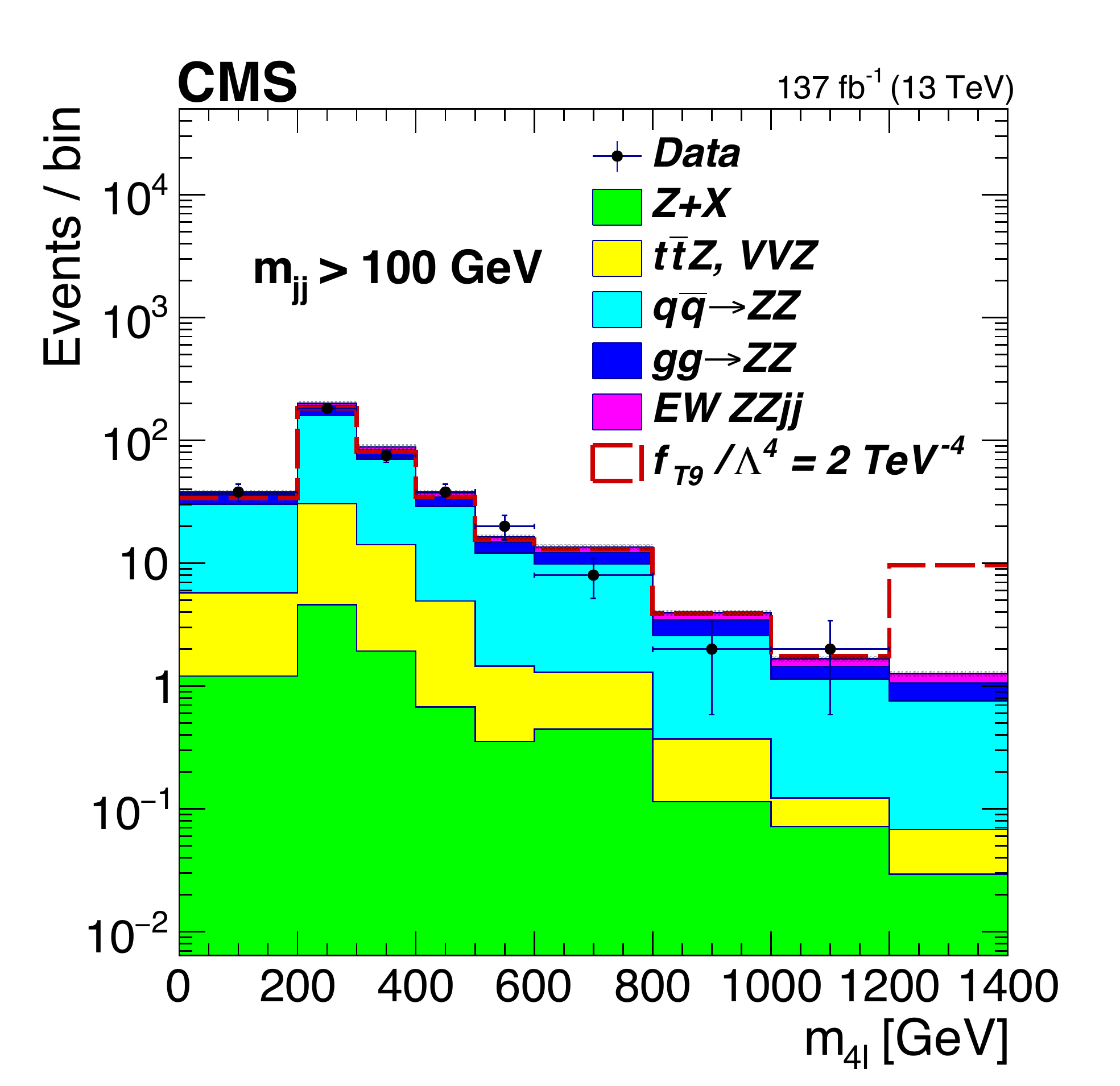}
\caption{Postfit distributions of the four-lepton invariant mass for  
$f_{\mathrm{T}9}/\Lambda^4$ and for events satisfying the \ZZjj inclusive selection. 
Points represent the data, filled histograms the fitted signal and background contributions, and the gray band the uncertainties derived from the fit covariance matrix. The expected distribution for an example value of $f_{\mathrm{T}9}/\Lambda^4 = 2\TeV^{-4}$ is also shown. The last bin includes all contributions with ${\mfourl>1200\GeV}$.}
\label{fig:atgc}
\end{figure}

Table~\ref{tab:xsec} reports the measured cross sections and their SM predictions in the three \ZZjj fiducial regions. 
For the SM predictions we report those extracted from generated events in MC samples adopted for the analysis, including the relative $K$ factors where applicable. For the EW \ZZjj prediction, in addition, we compare to 
higher-order calculations at NLO in QCD~\cite{Jager:2013iza,Alioli:2010xd} and with a theoretical prediction
at LO in QCD, but including NLO EW corrections~\cite{Denner:2020zit}. 
Uncertainties in all SM predictions come from variations of the factorization and renormalization scales. PDF$+\alpS$ variation uncertainties are summed in quadrature, except from the prediction from Ref.~\cite{Denner:2020zit} for which only the uncertainty in the scale variation is available.

The measured (expected) EW signal strength in the \ZZjj inclusive region is $\mu_{\mathrm{EW}} = 1.22^{+0.47}_{-0.40}$ ($1.00^{+0.44}_{-0.36}$). In the same region the measured (expected) EW+QCD signal strength is $\mu_{\mathrm{EW+QCD}} = 0.99^{+0.13}_{-0.12}$ ($1.00^{+0.13}_{-0.12}$). To quantify the significance of the EW signal, we compute the probability of the background-only hypothesis ($p$-value) as the tail integral of the test
statistic evaluated at $\mu_{\mathrm{EW}} = 0$ under the asymptotic approximation~\cite{CL_formula}. The background-only hypothesis is excluded with a significance of 4.0 (3.5 expected) standard deviations.

\begin{table*}[!t]
\centering
\topcaption{Measured cross sections and corresponding SM predictions in the three fiducial regions. The  reported SM predictions include those extracted from generated events in MC samples adopted for the analysis (LO), as well as
higher-order calculations at NLO in QCD~\cite{Jager:2013iza} (NLO QCD). }
   \begin{tabular}{l l c c}
   \hline 
	 & Perturbative order & SM $\sigma$ (fb) & Measured $\sigma$ (fb) \\
    \hline
	  \multicolumn{4}{c}{\ZZjj inclusive}\\ [\cmsTabSkip]
           \multirow{3}{*}{EW} & LO & $0.275 \pm 0.021$ &  
\multirow{3}{*}{$0.33^{+0.11}_{-0.10}\stat^{+0.04}_{-0.03}\syst$} \\                       
           & NLO QCD & $0.278 \pm 0.017$ & \\
           & NLO EW & $0.242^{+0.015}_{-0.013}$ & \\
           EW+QCD & & $5.35 \pm 0.51$ & $5.29 ^{+0.31}_{-0.30}\stat \pm 0.47\syst$\\  [\cmsTabSkip]
 
     \multicolumn{4}{c}{VBS-enriched (loose)}  \\ [\cmsTabSkip]
           \multirow{2}{*}{EW} & LO & $0.186 \pm 0.015$ & \multirow{2}{*}{$0.180^{+0.070}_{-0.060}\stat^{+0.021}_{-0.012}\syst$} \\
           & NLO QCD & $0.197 \pm 0.013$ & \\
            EW+QCD & & $1.21 \pm 0.09$ & $1.00^{+0.12}_{-0.11}\stat \pm 0.07\syst$\\  [\cmsTabSkip]
    
	  \multicolumn{4}{c}{VBS-enriched (tight)}\\ [\cmsTabSkip]
           \multirow{2}{*}{EW} & LO & $0.104 \pm 0.008$ & \multirow{2}{*}{$0.09 ^{+0.04}_{-0.03}\stat \pm 0.02\syst$} \\
           & NLO QCD & $0.108 \pm 0.007$ &\\
 
           EW+QCD & & $0.221 \pm 0.014$ & $0.20^{+0.05}_{-0.04}\stat \pm 0.02\syst$ \\ 
\hline
   \end{tabular}
      \label{tab:xsec}
\end{table*}

\section{Limits on anomalous quartic gauge couplings}
\label{sec:AQGC}
In an effective field theory approach to physics beyond the Standard Model, dimension-8 operators stem from 
covariant derivatives of the Higgs doublet and from charged and neutral field strength tensors associated to gauge
bosons. The latter generate eight independent operators, corresponding to couplings of the transverse degrees 
of freedom (T$i$) of the gauge fields.
The \ZZjj channel is particularly sensitive to the charged-current operators T0, T1, and T2, as well as the neutral-current operators T8 and T9 \cite{Eboli:2006wa}. The \mfourl distribution is used to constrain the aQGC parameters $f_{\mathrm{T}i}/\Lambda^4$, corresponding to the Wilson coefficients of the aforementioned operators, under the hypothesis of absence of anomalies in triple gauge couplings.
 
Figure~\ref{fig:atgc} shows the expected \mfourl distributions in the \ZZjj inclusive region, with postfit normalizations for the SM and for an example aQGC scenario, as well as the observed distribution in the data. The expected yield enhancement at large values of \mfourl exhibits a quadratic dependence on the anomalous couplings, and a parabolic function is fitted to the per-mass bin yields,
allowing for an interpolation between the discrete coupling parameters of the simulated aQGC signals.
The statistical analysis employs the same methodology used for 
the signal strength, including the profiling of the systematic uncertainties. 
using two different approaches, the distributions of the SM processes, including the EW component, are either normalized to their measured values in the EW signal extraction (as discussed in Section~\ref{sec:VBS}) or to their expected values. 
The Wald Gaussian approximation and Wilks' theorem are used to derive $2\sigma$ confidence level (\CL) intervals on the aQGC parameters~\cite{Junk:1999kv, CLS1,  CL_formula}. 
The measurement is statistically limited. 

\begin{table*}[!th]
\centering
\topcaption{Expected and observed limits of the 2$\sigma$ \CL intervals on the couplings of the quartic operators T0, T1, and T2, as well as the neutral current operators T8 and T9. Observed limits in parentheses are obtained by using the prefit normalization of SM processes. The unitarity bounds are also listed. All coupling parameter limits are in \TeVns{}$^{-4}$, while the unitarity bounds are in TeV.}
\begin{tabular}{l c c c c c}
\hline
Coupling		 	 &  Exp. lower	& Exp. upper	& Obs.  lower	& Obs. upper	& Unitarity bound\\
\hline
$f_{\mathrm{T}0}/\Lambda^4$         & $-0.37$  & 0.35  & $-0.24$ ($-0.26$)  & 0.22 (0.24)  & 2.4 \\
$f_{\mathrm{T}1}/\Lambda^4$         & $-0.49$  & 0.49  & $-0.31$ ($-0.34$)  & 0.31 (0.34) & 2.6 \\
$f_{\mathrm{T}2}/\Lambda^4$         & $-0.98$  & 0.95  & $-0.63$ ($-0.69$)  & 0.59 (0.65) & 2.5 \\
$f_{\mathrm{T}8}/\Lambda^4$         & $-0.68$  & 0.68  & $-0.43$ ($-0.47$)  & 0.43 (0.48) & 1.8 \\
$f_{\mathrm{T}9}/\Lambda^4$         & $-1.5$   & 1.5   & $-0.92$ ($-1.02$)  & 0.92 (1.02) & 1.8 \\
\hline
\end{tabular}
\label{tab:aqgc_limits}
\end{table*}

Table~\ref{tab:aqgc_limits} lists the individual lower and upper limits obtained by setting all other anomalous couplings to zero. The unitarity bounds 
are determined using the 
results from Ref.~\cite{Almeida:2020ylr}
as the scattering energy \mfourl at which the aQGC strength set equal to the observed limit would result in a scattering amplitude that violates unitarity.

\section{Summary}
\label{sec:summary}

A search was performed for the electroweak production of two jets in association with two \PZ bosons in the four-lepton final
state in proton-proton collisions at {13\TeV}.
The data correspond to an integrated luminosity of 137\fbinv 
collected with the CMS detector at the LHC.

The electroweak production of two jets in association with a pair of \PZ bosons
is measured with an observed (expected) significance of 4.0 (3.5) standard
deviations. The measured fiducial cross section is $\sigma_{\mathrm{fid}} = 0.33 ^{+0.11}_{-0.10}\stat^{+0.04}_{-0.03}\syst\unit{fb}$, which is consistent with the standard model prediction of $0.275 \pm 0.021\unit{fb}$.

Limits on anomalous quartic gauge couplings are set at 95\% confidence level in terms of effective field theory operators, with units in \TeVns{}$^{-4}$: 
\begin{center}
$-0.24 < f_{\mathrm{T}0}/\Lambda^4 < 0.22$ \\
$-0.31 < f_{\mathrm{T}1}/\Lambda^4 < 0.31$ \\
$-0.63 < f_{\mathrm{T}2}/\Lambda^4 < 0.59$ \\
$-0.43 < f_{\mathrm{T}8}/\Lambda^4 < 0.43$ \\
$-0.92 < f_{\mathrm{T}9}/\Lambda^4 < 0.92$ \\
\end{center}

These are the most stringent limits to date on the neutral current operators T8 and T9.

\begin{acknowledgments}

We congratulate our colleagues in the CERN accelerator departments for the excellent performance of the LHC and thank the technical and administrative staffs at CERN and at other CMS institutes for their contributions to the success of the CMS effort. In addition, we gratefully acknowledge the computing centres and personnel of the Worldwide LHC Computing Grid for delivering so effectively the computing infrastructure essential to our analyses. Finally, we acknowledge the enduring support for the construction and operation of the LHC and the CMS detector provided by the following funding agencies: BMBWF and FWF (Austria); FNRS and FWO (Belgium); CNPq, CAPES, FAPERJ, FAPERGS, and FAPESP (Brazil); MES (Bulgaria); CERN; CAS, MoST, and NSFC (China); COLCIENCIAS (Colombia); MSES and CSF (Croatia); RPF (Cyprus); SENESCYT (Ecuador); MoER, ERC IUT, PUT and ERDF (Estonia); Academy of Finland, MEC, and HIP (Finland); CEA and CNRS/IN2P3 (France); BMBF, DFG, and HGF (Germany); GSRT (Greece); NKFIA (Hungary); DAE and DST (India); IPM (Iran); SFI (Ireland); INFN (Italy); MSIP and NRF (Republic of Korea); MES (Latvia); LAS (Lithuania); MOE and UM (Malaysia); BUAP, CINVESTAV, CONACYT, LNS, SEP, and UASLP-FAI (Mexico); MOS (Montenegro); MBIE (New Zealand); PAEC (Pakistan); MSHE and NSC (Poland); FCT (Portugal); JINR (Dubna); MON, RosAtom, RAS, RFBR, and NRC KI (Russia); MESTD (Serbia); SEIDI, CPAN, PCTI, and FEDER (Spain); MOSTR (Sri Lanka); Swiss Funding Agencies (Switzerland); MST (Taipei); ThEPCenter, IPST, STAR, and NSTDA (Thailand); TUBITAK and TAEK (Turkey); NASU (Ukraine); STFC (United Kingdom); DOE and NSF (USA). 

\hyphenation{Rachada-pisek} Individuals have received support from the Marie-Curie programme and the European Research Council and Horizon 2020 Grant, contract Nos.\ 675440, 752730, and 765710 (European Union); the Leventis Foundation; the A.P.\ Sloan Foundation; the Alexander von Humboldt Foundation; the Belgian Federal Science Policy Office; the Fonds pour la Formation \`a la Recherche dans l'Industrie et dans l'Agriculture (FRIA-Belgium); the Agentschap voor Innovatie door Wetenschap en Technologie (IWT-Belgium); the F.R.S.-FNRS and FWO (Belgium) under the ``Excellence of Science -- EOS" -- be.h project n.\ 30820817; the Beijing Municipal Science \& Technology Commission, No. Z191100007219010; the Ministry of Education, Youth and Sports (MEYS) of the Czech Republic; the Deutsche Forschungsgemeinschaft (DFG) under Germany's Excellence Strategy -- EXC 2121 ``Quantum Universe" -- 390833306; the Lend\"ulet (``Momentum") Programme and the J\'anos Bolyai Research Scholarship of the Hungarian Academy of Sciences, the New National Excellence Program \'UNKP, the NKFIA research grants 123842, 123959, 124845, 124850, 125105, 128713, 128786, and 129058 (Hungary); the Council of Science and Industrial Research, India; the Italian and Serbian Ministries for Foreign Affairs and International Cooperation (MAECI/MFA), grant n.\ RS19MO06 (Italy-Serbia); the HOMING PLUS programme of the Foundation for Polish Science, cofinanced from European Union, Regional Development Fund, the Mobility Plus programme of the Ministry of Science and Higher Education, the National Science Center (Poland), contracts Harmonia 2014/14/M/ST2/00428, Opus 2014/13/B/ST2/02543, 2014/15/B/ST2/03998, and 2015/19/B/ST2/02861, Sonata-bis 2012/07/E/ST2/01406; the National Priorities Research Program by Qatar National Research Fund; the Ministry of Science and Education, grant no. 14.W03.31.0026 (Russia); the Tomsk Polytechnic University Competitiveness Enhancement Program and ``Nauka" Project FSWW-2020-0008 (Russia); the Programa Estatal de Fomento de la Investigaci{\'o}n Cient{\'i}fica y T{\'e}cnica de Excelencia Mar\'{\i}a de Maeztu, grant MDM-2015-0509 and the Programa Severo Ochoa del Principado de Asturias; the Thalis and Aristeia programmes cofinanced by EU-ESF and the Greek NSRF; the Rachadapisek Sompot Fund for Postdoctoral Fellowship, Chulalongkorn University and the Chulalongkorn Academic into Its 2nd Century Project Advancement Project (Thailand); the Kavli Foundation; the Nvidia Corporation; the SuperMicro Corporation; the Welch Foundation, contract C-1845; and the Weston Havens Foundation (USA). 

\end{acknowledgments}

\bibliography{auto_generated} 
\cleardoublepage \appendix\section{The CMS Collaboration \label{app:collab}}\begin{sloppypar}\hyphenpenalty=5000\widowpenalty=500\clubpenalty=5000\vskip\cmsinstskip
\textbf{Yerevan Physics Institute, Yerevan, Armenia}\\*[0pt]
A.M.~Sirunyan$^{\textrm{\dag}}$, A.~Tumasyan
\vskip\cmsinstskip
\textbf{Institut f\"{u}r Hochenergiephysik, Wien, Austria}\\*[0pt]
W.~Adam, F.~Ambrogi, T.~Bergauer, M.~Dragicevic, J.~Er\"{o}, A.~Escalante~Del~Valle, R.~Fr\"{u}hwirth\cmsAuthorMark{1}, M.~Jeitler\cmsAuthorMark{1}, N.~Krammer, L.~Lechner, D.~Liko, T.~Madlener, I.~Mikulec, F.M.~Pitters, N.~Rad, J.~Schieck\cmsAuthorMark{1}, R.~Sch\"{o}fbeck, M.~Spanring, S.~Templ, W.~Waltenberger, C.-E.~Wulz\cmsAuthorMark{1}, M.~Zarucki
% ...
\vskip\cmsinstskip
\dag: Deceased\\
1:  Also at Vienna University of Technology, Vienna, Austria\\
% ...
\end{sloppypar}
%%% END EDITABLE REGION %%%
\end{document}